\documentclass[journal=jacsat,manuscript=article]{achemso}

\usepackage{mathpazo}
\usepackage{microtype}
\usepackage{float}
\usepackage{geometry}
\usepackage{natbib}
\usepackage{setspace}
\usepackage{xkeyval}
\usepackage{chemformula}
\usepackage{xcolor}
\makeatletter
\newcommand*{\rom}[1]{\expandafter\@slowromancap\romannumeral #1@} 

\author{Kati Asikainen}
\email{Kati.Asikainen@oulu.fi}
\author{Matti Alatalo}
\author{Marko Huttula}
\affiliation[lab1]
{Nano and Molecular Systems Research Unit, University of Oulu, FI-90014, Finland}
\author{Assa Aravindh Sasikala Devi}
\email{Assa.SasikalaDevi@oulu.fi}
\affiliation[lab2]
{Research Unit of Sustainable Chemistry, University of Oulu, FI-90014, Finland}

\title{Data-Efficient Machine learning for Predicting Dopant Formation Energies in \ch{TiO2} Monolayer}

\begin{document}
\clearpage

\begin{abstract}

Machine learning models are increasingly applied in materials science, yet their predictive power is often constrained by data scarcity. Here, we show that accurate predictions can be achieved, even with a limited number of training examples, provided the dataset is compact and and grounded in physically relevant quantities. By combining density functional theory calculations with a machine-learning framework, we construct accurate descriptor-based models to predict the formation energies of doped lepidocrocite \ch{TiO2} monolayers. The predictive accuracy of machine-learning models was first evaluated for single-dopant Pt configurations, demonstrating that the selected structural and chemical descriptors reliably capture the key factors governing dopant stability. Chemical transferability is then examined by extending the dataset to include Ag-doped configurations. Predictive accuracy improved systematically as additional Ag-doped data points were included in the training, while the performance of Pt remains robust. These results highlight the potential of small and well-curated datasets combined with physically informed descriptors to enable not only accurate but also chemically transferable machine-learning-driven screening. 

\end{abstract}

\section{Introduction}

Two-dimensional (2D) materials have attracted significant research interest in material science since the exfoliation of graphene for graphite in 2004 \cite{graphene}. Their atomic-scale thickness gives rise to pronounced quantum confinement effects, resulting in physical and chemical properties distinct from those of their bulk counterparts. This has enabled diverse applications, including but not limited to catalysis, energy storage, and biomedical applications \cite{2D-1, 2D-2}. Beyond symmetric 2D materials, designing Janus structures - characterized by out-of-plane asymmetry withing a single 2D crystal - offers an additional degree of freedom to tailor material properties \cite{Janus, JanusRev}. However, because pristine materials often exhibit limited performance for practical applications, doping remains a widely used strategy to tune material properties. Doping introduces defect states, leading to modifications in the band structure, charge distribution, and chemical reactivity. While density functional theory (DFT) has been extensively employed to unravel atomic-level structural and electronic changes in doped 2D materials \cite{Rahman2025,Pacchioni2008,Freysoldt2014}, machine learning (ML) approaches have recently emerged as powerful tools to accelerate scientific progress beyond conventional simulations \cite{ML-1,ML-2}. 

Among transition metal oxides, titanium dioxide (\ch{TiO2}) is a benchmark material in photocatalysis due to its chemical stability, low toxicity, and broad technological relevance. Its 2D lepidocrocite phase, first synthesized by Sasaki and Watanabe \cite{LNS}, has attracted increasing attention in recent years. Unfortunately, similar to bulk \ch{TiO2}, it suffers performance degradation due to limited light adsorption and rapid charge recombination. However, numerous studies have demonstrated that 2D \ch{TiO2} offers favorable structural and electronic tunability through doping \cite{Jin2019,Kim2017,Yuan2019,Pradhan2023,Li2020,Yuan2020,Reeves2018,JanusTiO2}. Noble metal dopants, in particular, remain of strong interest due to their superior catalytic efficiency and selectivity, although their high cost and scarcity pose challenges in large-scale deployment. In substitutional doping, dopant atoms most often replace Ti sites in the \ch{TiO2} lattice, but incorporation at O sites is also possible. We have recently adapted this strategy and designed Janus monolayers based on 2D \ch{TiO2} by substitutionally doping one surface oxygen layer with noble metals (Ag, Au, Pd, and Pt) \cite{JanusTiO2}. 
This structural asymmetry resulted in an enhanced water interaction and optimized hydrogen adsorption, reaching a performance comparable to that of the Pt(111) surface.

Although DFT provides valuable atomic-scale insight, such studies are typically limited to a small number of configurations. The high computational costs make it impractical to systematically explore the vast configurational space of dopant types, positions, and concentrations. Moreover, doping can alter not only the properties, but also the stability of 2D materials, and understanding these effects becomes increasingly challenging given the combinatorial number of possible configurations. These limitations motivate the use of ML approaches to develop predictive models that enable rapid screening of doped materials. When in comes to \ch{TiO2}, ML has been applied to study oxygen vacancy formation \cite{Wang2024} and their interaction with polarons \cite{Birschitzky2024}, as well as to optimize single-atom dopant selection at the Ti site in bulk \ch{TiO2} \cite{Jiang2024}. Gao \textit{et al.} \cite{Gao2021} further demonstrated the ML-driven identification of optimal synthesis parameters for C/N co-doped \ch{TiO2} nanoparticles. In addition, several studies have integrated ML to predict key properties of doped \ch{TiO2}, including band gap \cite{Zhang2020,Kurban2024}, photocatalytic mechanisms and performance \cite{Mikolajczyk2024, Liu2022, Azadi2018, Jiang2020, Liu2023, Ayodele2020, Saleem2025, Meng2025} and battery-related behavior \cite{Jiang2023}. However, ML investigations on 2D lepidocrocite \ch{TiO2} remain scarce. This gap presents an opportunity to combine DFT and ML to extend data-driven approaches and systematically explore the factors that govern the stability of doped 2D \ch{TiO2}. Notably, the formation energy has been identified as a key descriptor in modeling HER performance \cite{Jyothirmai2023,Tamtaji2022}, underscoring the relevance of understanding formation energetics. 

While ML models often rely on large datasets to achieve high predictive accuracy, such data volumes are not always accessible in materials science. In this context, a more realistic objective is to construct compact, carefully selected datasets that still enable accurate and meaningful predictions. Building on our previous work \cite{JanusTiO2}, DFT calculations are combined with a ML framework to identify structural and chemical descriptors governing the stability of doped 2D \ch{TiO2}. Here, we deliberately restrict the analysis to a small database to evaluate the data efficiency of the proposed approach in predicting the formation energies. The study proceeds in two stages. First, ML performance is assessed using Pt-doped configurations, and descriptors that best capture the dependence of formation energy on dopant concentration and local structure. The chemical transferability is examined by extending the analysis to Ag-doped configurations. Cross-validation is applied throughout to assess model robustness and to evaluate whether the available dataset size and composition are adequate to support stable ML performance. 


\section{Methodology}

\subsection{Density functional theory}
First-principles density functional theory (DFT) calculations were performed using the Vienna Ab initio Simulation package (VASP) code 6 \cite{vasp1, vasp2, vasp3}. The projected augmented wave (PAW) pseudopotentials \cite{PAW} were used to describe the ion-electron interaction and the generalized gradient approximation (GGA) according to Perdew-Burke-Ernzerhof (PBE) exchange-correlation functional \cite{GGA}. The plane-wave energy cutoff was set to 520 eV. Following the same methodology as in our previous study \cite{JanusTiO2}, the GGA was employed for geometry optimization due to its reliability in predicting structural parameters. Because the lattice vectors experienced negligible dopant-induced stress in a large supercell, only ionic relaxation was allowed in the calculations. The electronic properties were then calculated using the GGA+U approach according to the Dudarev scheme \cite{GGA+U} using $U_{Ti}$ = 4.5 eV and $U_{Pt}=$ 2.5 eV. The k-point grid of $2 \times 2 \times 1$ according to the Monkhorst-Pack scheme \cite{MP} was used to sample the Brillouin zone. Electronic occupancies were treated using the Gaussian smearing method with a smearing width of 0.05 eV. The convergence criterion for energy and forces was set to $10^{-6}$ eV and 0.001 eV/\AA \ in all calculations, respectively. Dipole correction was included when calculating the electrostatic potential. The post-processing of the VASP output was done using the VESTA \cite{VESTA}, VASPKIT \cite{vaspkit}, and Bader Charge Analysis by Henkelman's group \cite{bader1, bader2}. 

The formation energy of the doped configurations per Pt atom, which serves as the target variable in ML models, was calculated using the formula
\begin{align}
E_{\text{formation}} = \frac{E_{\text{doped}} - E_{\text{undoped}} + n\mu_{\text{O}} - n\mu_{\text{dopant}}}{n}
\end{align}
where $E_{\text{doped}}$ and $E_{\text{undoped}}$ are the total energies of the doped and pristine \ch{TiO2} monolayer, $\mu_{\text{O}}$ and $\mu_{\text{dopant}}$ are the chemical potentials of O and noble metal dopant (Pt or Ag), and $n$ is the number of dopant atoms. The chemical potentials of O and noble metals are derived from an isolated \ch{O2} molecule and cubic bulk structures, respectively.

\subsection{Machine learning analysis}
Nine ML models were benchmarked, including Linear Regression (LR), Lasso Regression (LASSO), Ridge Linear Regression (RLR), Elastic Net (EN), Random Forest Regression (RF), Gradient Boosting Regression (GBR), k-Nearest Neighbors Regression (KNN), Support Vector Regression (SVR), and Gaussian Process Regression (GPR), which were implemented using the open-source \textit{Scikit-learn} python code \cite{scikit}. These models are widely used in material science, including for 2D materials \cite{Wei2019,Lu2024}. To assess the predictive performance of the models, the coefficient of determination ($R{}^2$), the root-mean-square error (RMSE), and the mean absolute error (MAE) were calculated. These evaluation metrics are defined as follows: 
\begin{align}
R{}^2 = 1-\frac{\sum_{i=1}^n (y_i-\hat{y})²}{\sum_{i=1}^n (y_i-\bar{y})²}, \label{R}\\[1em] 
\text{RMSE} = \sqrt{\frac{1}{n}\sum_{i=1}^n (y_i-\hat{y})²}, \\[1em]
\text{MAE} = \frac{1}{n}\sum_{i=1}^n |y_i-\hat{y}|,
\end{align}
where $y_i$ denotes the DFT-calculated value, $\hat{y}$ represents the ML-predicted value, $\bar{y}$ is the mean of the DFT-calculated values, and $n$ corresponds to the number of data points used in the regression model. The hyperparameters of the ML models were optimized using a 5-fold cross validation (CV). Additionally, the evaluation metrics defined above were recomputed through cross-validation to assess the models' robustness and generalization of the models, which is essential when working with relatively small datasets.

\section{Results}

\subsection{Dataset compilation and feature extraction for Pt-doped monolayers}

\begin{figure}[h!]\centering
\includegraphics[width=1\linewidth]{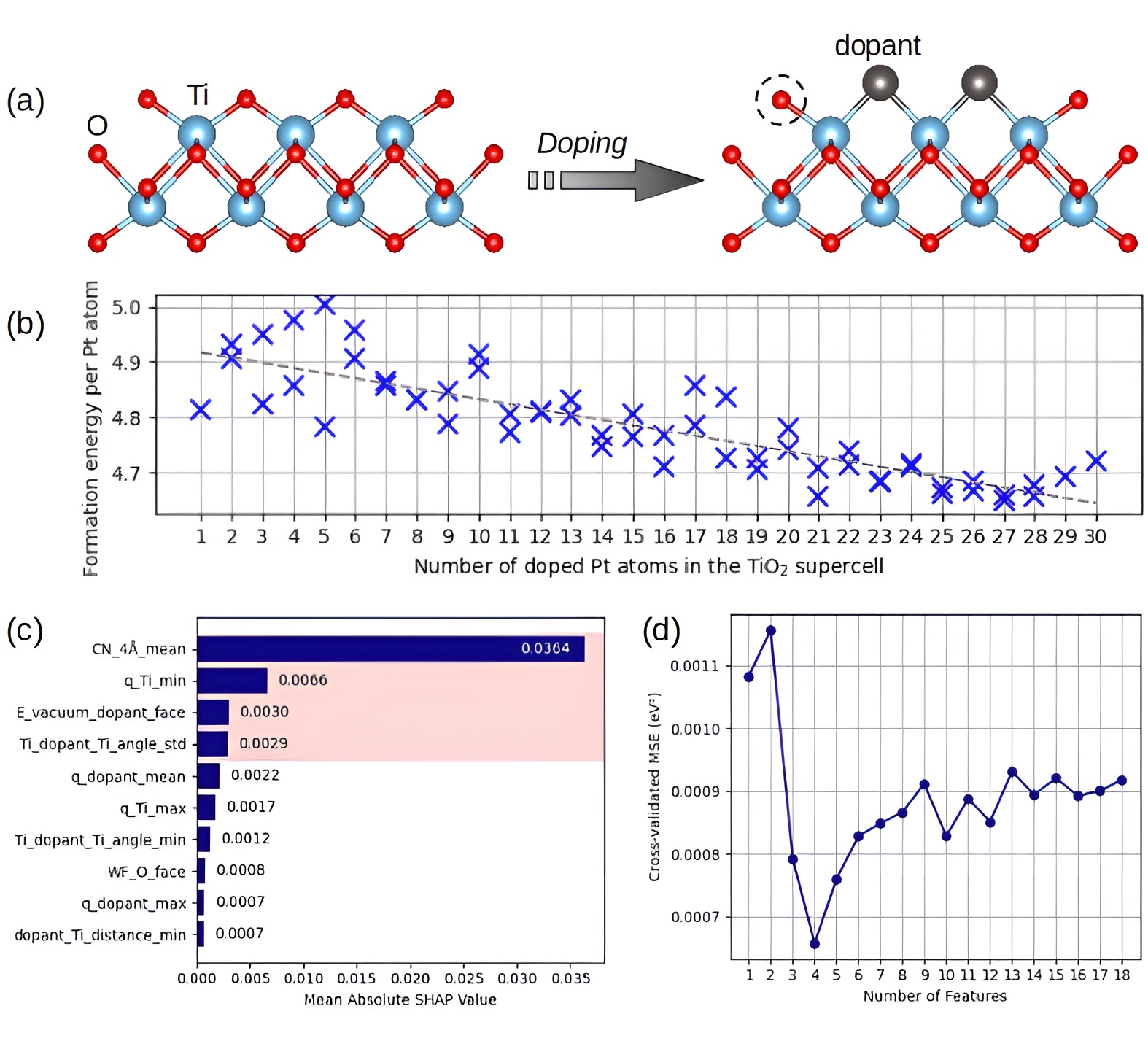}
\caption{(a) Schematic representation of substitutional doping at the $O_b$ site in a \ch{TiO2} monolayer. (b) Per-atom formation energy of Pt-doped configurations plotted against the number of doped Pt atoms. The gray dashed line shows the least-squares fit, indicating an inverse relationship between the variables. (c) Mean absolute SHAP value of the most importance features (subset shown). The shaded coral region highlights the four features selected for ML predictions based on the recursive feature elimination of Pt-doped dataset, shown in (d).}\label{Fig1}
\end{figure}

We constructed a $6 \times 5 \times 1$ supercell of the relaxed 2D lepidocrocite \ch{TiO2} monolayer, comprising a total of 180 atoms (60 Ti and 120 O atoms), to generate the dataset. Substitutional doping was then carried out at the two-fold coordinated bridging oxygen ($O_b$) sites on one side of the monolayer (Fig. \ref{Fig1}a). Each surface contains 30 $O_b$ sites, allowing dopant concentrations ranging from $N=1$ to complete substitution ($N=30$). The fully substituted Janus configuration was previously investigated \cite{JanusTiO2}. Among the four dopants considered (Ag, Au, Pd, and Pt), the Pt-doped monolayer exhibited the lowest formation energy and was therefore selected as the reference system. For $N=1$, 29, and 30, only one symmetry-distinct configuration exists, and therefore they were placed directly in the test set, allowing for a natural assessment of the performance of the models. For the remaining concentrations, two symmetry-distinct configurations were generated and randomly split into the training set and the test set using an 80/20 split. In total, the dataset comprises 57 Pt-doped configurations, with 44 used for training and 13 for testing. To ensure physical relevance, only configurations with formation energies below 6 eV were retained. 

DFT calculations were performed for each doped configuration to generate input data for the ML models. Fig. \ref{Fig1}b shows the formation energies per Pt for configurations containing different numbers of Pt dopants. A least-squares fit indicates an inverse trend: the formation energy decreases with increasing dopant concentration. The maximum variation is 0.36 eV per atom, indicating that dopant incorporation becomes progressively easier. To probe structure-property relationships, structural (Fig. S1) and chemical descriptors were extracted from the relaxed doped configurations, forming the initial ML feature space (Table S1). Because the number of variables increases with the dopant count, the variable-length descriptors were converted into fixed-length representations using statistics measures (minimum, maximum, mean, and standard deviation). To ensure a manageable number of features relative to the datasets, feature-selection methods were applied to identify the most informative descriptors and to reduce the feature space.   

Linear feature relationships were examined using the Pearson correlation analysis \cite{}, followed by feature-importance analysis based on mean absolute SHAP \cite{shap} values to capture non-linear dependencies (Fig. S2). To mitigate multicollinearity, pairs of strongly correlated features ($|p| > 0.9$) were identified, and in each pair the feature with the lower SHAP importance (from an RF model) was removed. The SHAP importance of the most relevant screened features is shown in Fig. \ref{Fig1}c. Notably, the dominant feature, the average coordination number within a 4\ \AA \  radius of a dopant atom ($\text{CN}_{-}\text{4Å}_{-}\text{mean}$), substantially drives the model predictions, exhibiting an importance nearly five times higher than the second-ranked descriptor. 

To identify the optimal number of input features for the ML predictions, recursive model evaluation was performed using cross-validation while adding features in order of decreasing importance. Based on the mean square error (MSE), the performance reaches a global minimum at four features (Fig. \ref{Fig1}d), indicating that this subset provides the best predictive capacity. The following four descriptors were selected for the ML stage: $\text{CN}_{-}\text{4Å}_{-}\text{mean}$ (average coordination number of a dopant atom within a 4\ \AA \  radius), $\text{q}_{-}\text{Ti}_{-}\text{min}$ (minimum Bader charge of the Ti atoms nearest to dopant atoms), $\text{E}_{-}\text{vacuum}_{-}\text{dopant}_{-}\text{face}$ (vacuum level of the dopant-doped monolayer surface), $\text{Ti}_{-}\text{dopant}_{-}\text{Ti}_{-}\text{angle}_{-}\text{std}$ (standard deviation of all Ti-dopant-Ti angles) (Fig. \ref{Fig1}c). 

The SHAP summary plot (Fig. S3) reveals an inverse relationship between the formation energy and most input features, with the exception of  $\text{q}_{-}\text{Ti}_{-}\text{min}$. In general, less negative $\text{q}_{-}\text{Ti}_{-}\text{min}$ corresponds to lower dopant concentrations and contributes positively to the target value. The dominant descriptor, $\text{CN}_{-}\text{4Å}_{-}\text{mean}$, strengthens the negative contribution at higher dopant concentrations. This suggests that an increased number of neighboring dopant atoms lowers the formation energy, indicating that collective dopant environments are energetically more favorable. 

\subsection{Machine learning predictions for Pt-doped monolayers}

\begin{figure}[h!]\centering
\includegraphics[width=0.9\linewidth]{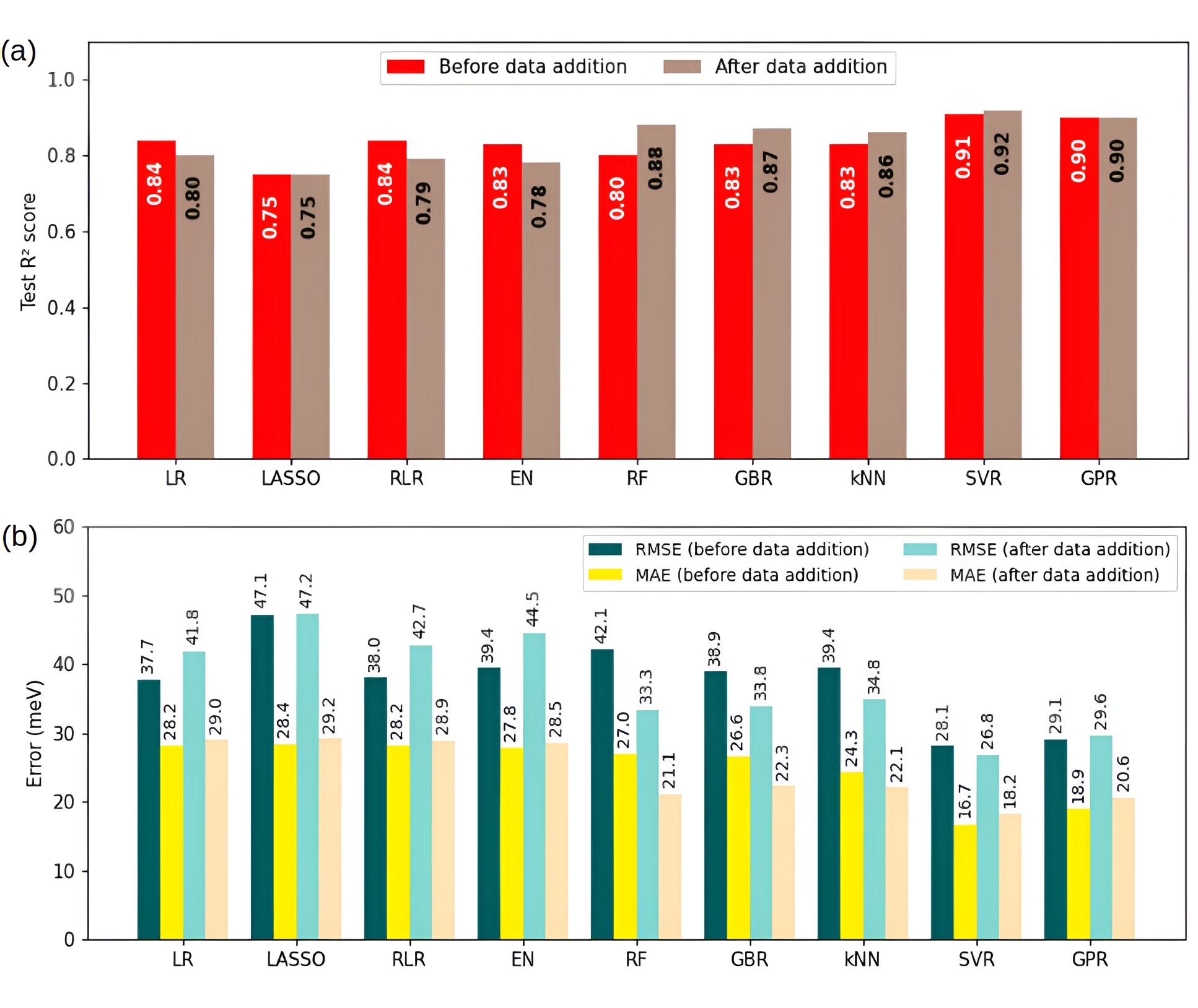}
\caption{(a) Test $R{}^2$ for Pt-doped dataset and (b) corresponding RMSE and MAE values (meV per Pt atom) before and after data addition.}\label{Fig2}
\end{figure}

\begin{figure}[h!]\centering
\includegraphics[width=1\linewidth]{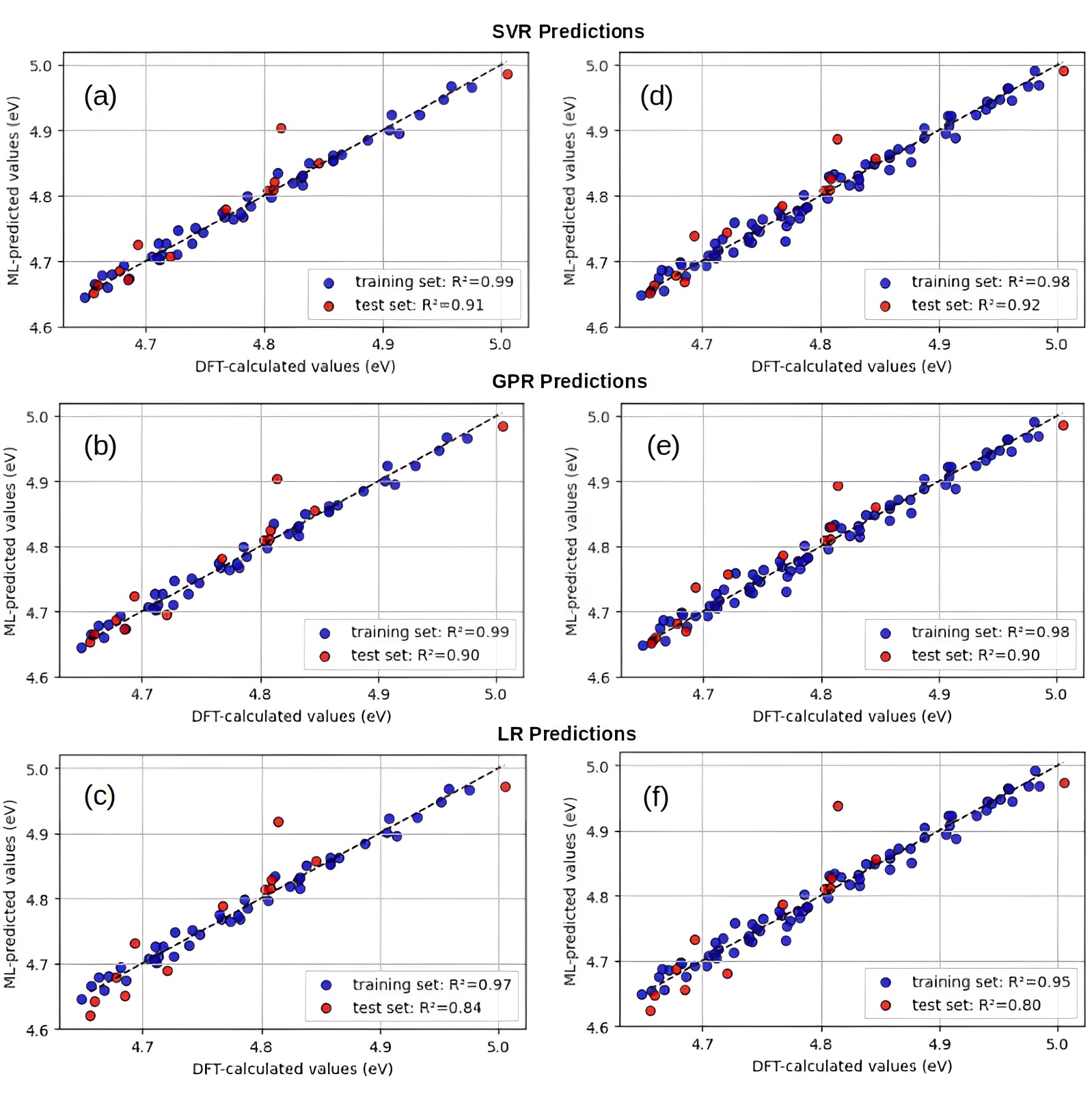}
\caption{DFT-calculated versus ML-predicted formation energy per Pt atom in eV for the training and test sets using (a,d) SVR, (b,e) GPR, and (c,f) LR. Left column panels (a-c) show the results using the original training set, and the right column panels (d-f) the results after expanding the training set. Black dashed line corresponds to the perfect predictions.}\label{Fig3}
\end{figure}

With the optimal input features identified, nine ML models (listed in the Methodology section) were trained to predict the formation energy per Pt atom. After hyperparameter optimization, all models achieved the test $R{}^2$ between 0.75-0.91, accompanied by low RMSE and MAE values below 50 meV (Fig. \ref{Fig2}). These results indicate that the selected feature subset sufficiently captures the underlying trends of the dataset. The top-performing models were SVR, GPR, and LR, reaching a test $R{}^2$ of 0.91, 0.90, and 0.84, respectively. Their prediction results are shown in Fig. \ref{Fig3}a-c. The corresponding RMSE values were 28.1, 29.1 and 37.7 meV per Pt atom, and the MAE values 16.7, 18.9 and 28.2 meV per Pt atom, respectively. 

Because a small dataset can limit model robustness, the original training set was expanded with additional data points, and the impact on models' performance was assessed. For each $N=2-28$, a new, previously unconsidered doped configuration was generated, increasing the training set from 44 to 71 data points. The input features were kept unchanged, while the hyperparameters were re-optimized during retraining. The results show that the test $R{}^2$ remains largely unchanged across all models when using the expanded dataset (Fig. \ref{Fig2}). The variation in $R{}^2$ is generally $\le$ 0.05, and the error metrics also remain similar. This indicates that the added data had little impact on model performance, suggesting that the original dataset is already representative enough for effective model training. Training $R{}^2$ remained consistently high ($\ge  0.94$), supporting stable model fits.

The cross-validation results initially suggest a slightly optimistic estimate of model performance (Fig. S4a). However, as the training set expanded, the CV $R{}^2$ converged toward the test $R{}^2$, indicating that a larger training dataset yields a less biased and more realistic assessment. CV RMSE and MAE remain slightly lower than their test-set counterparts (Fig. S4b), even after adding data, implying a minor degree of overfitting, in line with the observed gap between the training and test $R{}^2$ scores. Nevertheless, the differences are small - RMSE $\le$ 25 meV and MAE $\le$ 10 meV per Pt atom - relative to the scale of the target property, suggesting the overall reliability of the models. 

The best-performing models achieve a test $R{}^2 \approx 0.9$, capturing most of the variance in the formation energy. A single outlier-like point was observed appearing consistently across all models (Fig. \ref{Fig3}, S5 and S6), corresponding to the unique $N=1$ configuration with a formation energy of 4.81 eV. As the calculations successfully converged within the specified tolerance limits, this deviation is attributed to the underrepresentation of its distinct chemical environments. Because $N=1$ is unique, even an expanded dataset may provide little opportunity for models to learn from. In contrast, the other edge cases $N=29$ and $N=30$ are better captured due to their greater similarity to the dataset. Some apparent outliers arise from model-specific inductive biases rather than deficiencies in the data. Nevertheless, we retained all data points in the analysis to preserve the physical diversity of the dataset.

It should be noted that the study employs a set of physically and chemically motivated descriptors, similar to those widely used in the literature \cite{Jiang2024, Pentyala2022, Choudhary2018,Cheng2020,Zhang2021,Jyothirmai2023,Tamtaji2022}, because the most informative descriptors were not known apriori and can be system-dependent. Although additional or different descriptors might further enhance predictive performance, identifying them typically requires an iterative process. Despite this, the models already perform strongly with the selected features, offering a solid baseline. 

\subsection{Assessing predictive transfer to Ag-doped monolayers}

Since the models were trained on Pt-doped monolayers, their transferability to chemically distinct elements remain uncertain. To examine this, the dataset was extended to include Ag-doped monolayers. Owing to their distinct formation energy range, Ag-doped configurations provide a clear out-of-distribution benchmark. Previous studies have demonstrated generalization across dopant species in \ch{TiO2} \cite{Jiang2024, Zhang2020, Mikolajczyk2024, Liu2023}, and have suggested that a small, representative subset of data from a distinct chemical domain may help align models with the new distribution and enhance generalization \cite{Harper2022, Rocken2025, Li2024}. 

Consequently, 14 representative Ag-doped configurations were selected spanning low, intermediate, and high dopant concentrations (more precisely, $N =$ 1, 2, 4, 5, 7, 9, 10, 13, 16, 18, 22, 24, 28, 29). In this testing setup, multiple configurations per dopant concentration were not considered. Compared to the Pt-doped dataset, the 14 Ag-doped configurations provide a small, targeted dataset to assess the models’ data-efficiency. This dataset comprises both configurations analogous to the Pt-doped systems and entirely new arrangements. 

\begin{figure}[h!]\centering
\includegraphics[width=1\linewidth]{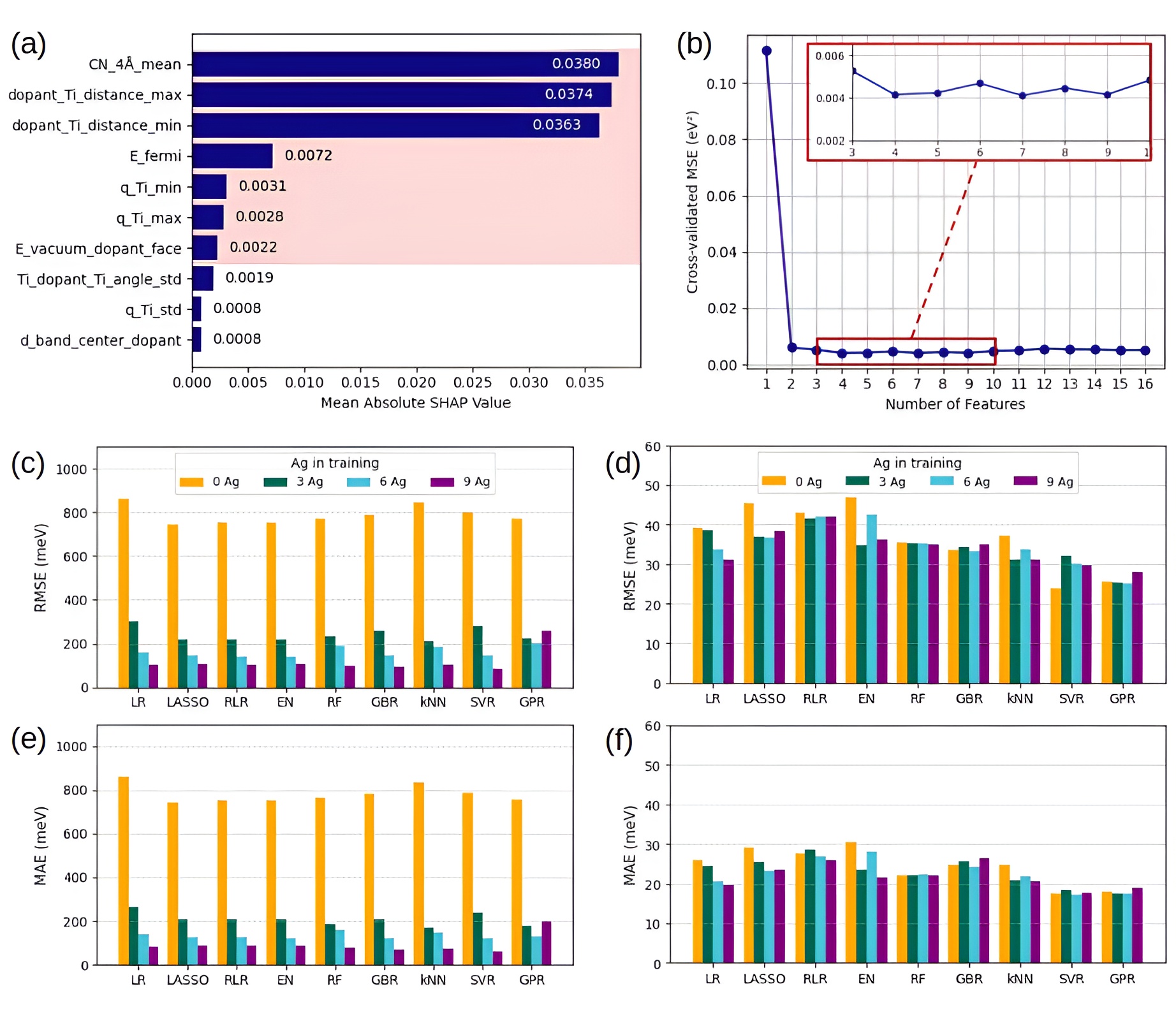}
\caption{(a) Mean absolute SHAP value of the most importance features (subset shown) after inclusion of Ag-doped data. The seven features selected for ML predictions based on the recursive feature elimination of Pt-doped dataset, shown in (b), are highlighted with a shaded coral region. (c,e) Per-element RSME and MAE of Ag-doped dataset and (d,f) the corresponding metrics of Pt-doped dataset versus the number of Ag-doped data points used for training.}\label{Fig4}
\end{figure}

We then repeated the feature-analysis procedure, feature importance is not necessarily invariant. To enable the models to distinguish between dopant species, an elemental identifier - the atomic number Z (Pt: 78, Ag: 47) - was added. Because it serves only as a label, it was excluded from the feature analysis. Pearson correlation analysis showed that the multicollinearity structure changes with the inclusion of Ag data, resulting in a different set of screened features (Fig. S7). Based on SHAP analysis and recursive feature elimination (Fig. \ref{Fig4}a,b), seven optimal descriptors were identified: $\text{CN}_{-}\text{4Å}_{-}\text{mean}$, $\text{dopant}_{-}\text{Ti}_{-}\text{distance}_{-}\text{max}$ (maximum dopant-Ti distance), $\text{dopant}_{-}\text{Ti}_{-}\text{distance}_{-}\text{min}$ (minimum dopant-Ti distance), $\text{E}_{-}\text{Fermi}$ (Fermi level), $\text{q}_{-}\text{Ti}_{-}\text{min}$ (minimum Bader charge of dopant atoms), $\text{q}_{-}\text{Ti}_{-}\text{max}$ (maximum Bader charge of dopant atoms), and $\text{E}_{-}\text{vacuum}_{-}\text{dopant}_{-}\text{face}$. The number of input features increased from four to seven (Fig. \ref{Fig1}c,d and Fig. \ref{Fig4}a,b) when moving from the Pt-doped monolayers to the combined dataset. Some descriptors remain largely dopant-independent (e.g. $\text{CN}_{-}\text{4Å}_{-}\text{mean}$), while others are dopant-sensitive, reflecting intrinsic differences between Pt and Ag-doped monolayers. This indicates that additional descriptors are needed to capture dopant-dependent variations in the formation energy. Moreover, with Ag included, feature importance becomes more evenly distributed. Three descriptors contribute comparably, while the remaining features act as moderate refinements (Fig. \ref{Fig4}a). $\text{CN}_{-}\text{4Å}_{-}\text{mean}$ maintains an inverse relationship with the formation energy, while the next most informative descriptors,  $\text{dopant}_{-}\text{Ti}_{-}\text{distance}_{-}\text{max}$ and $\text{dopant}_{-}\text{Ti}_{-}\text{distance}_{-}\text{min}$, exhibit nearly a binary-like influence on the predictions, with large values driving the formation energy upward (Fig. S8). 

To ensure the representation of both dopants in the training and test set, a stratified split was applied, with random sampling retained within each subset. The Pt-doped data were again divided using an 80:20 train-set split, this time without forcing extreme points into the test set. For the Ag-doped data, five configurations were randomly assigned to the test set, and the remaining points were progressively added to the training set. 
 
Models were first trained exclusively on Pt-doped data. While Pt-doped configurations are well described, the models completely fail to predict Ag-doped formation energies (Fig. S9). This highlights the challenge of chemical extrapolation beyond the dopant domain represented in the training data. 

\begin{figure}[h!]\centering
\includegraphics[width=1\linewidth]{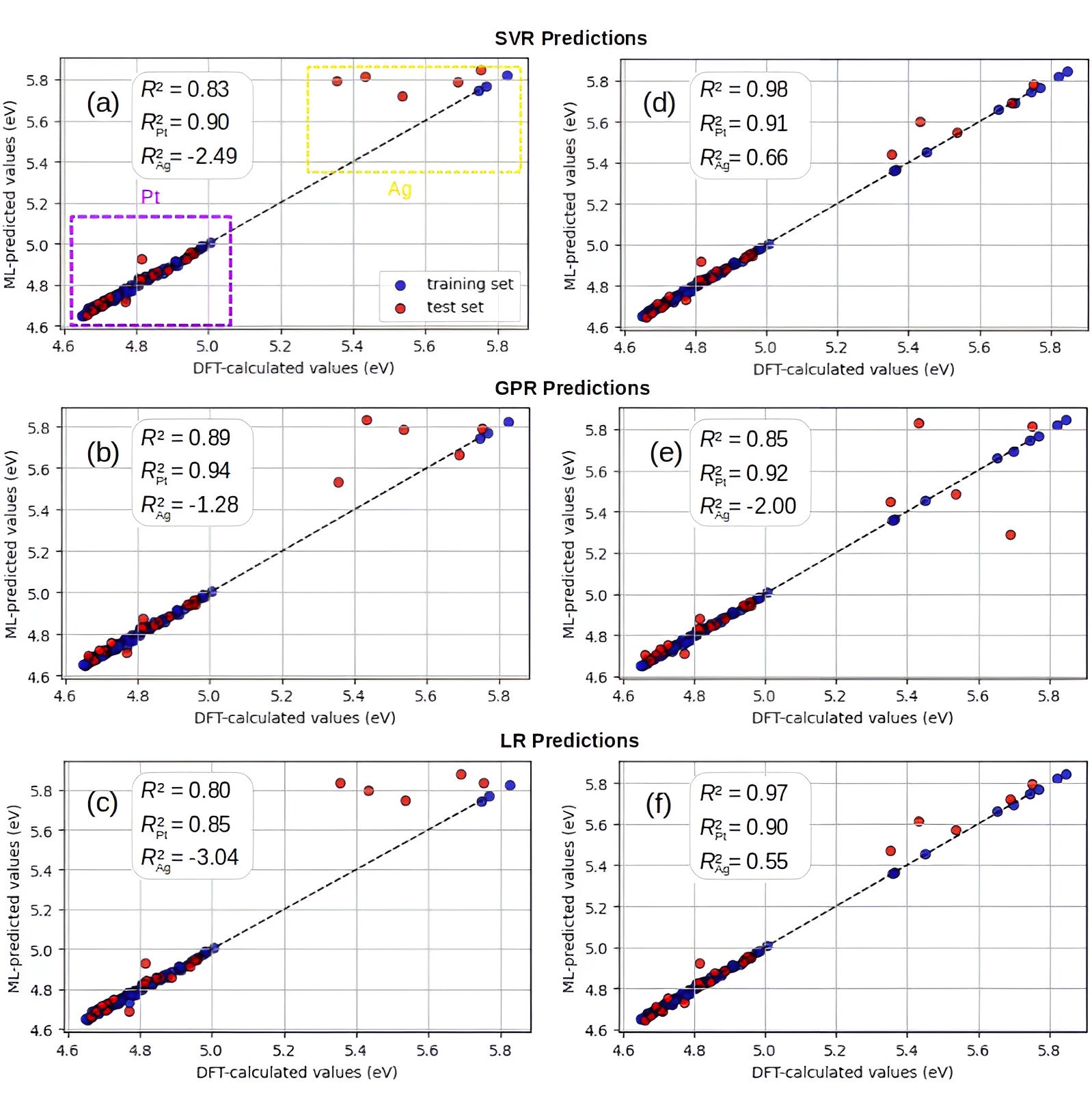}
\caption{DFT-calculated versus ML-predicted formation energy per Pt atom in eV for the training and test sets using (a,d) SVR, (b,e) GPR, and (c,f) LR. Left column panels (a-c) show the results after adding 3 Ag-doped data points, and the right column panels (d-f) after adding 9 Ag-doped data points to the training set. Black dashed line corresponds to the perfect predictions. Pt and Ag-doped data points are highlighted by purple and yellow dashed squares in the top left plot, respectively.}\label{Fig5}
\end{figure}

The models were then trained by progressively adding Ag data to the training set, rapidly learning to capture chemically distinct behavior. Even a small number of Ag-doped configurations enabled recognition of dopant-specific trends, and additional Ag data steadily improved predictive performance (Fig. \ref{Fig4}c,e and Table S2 and S4), as further illustrated by DFT-versus-ML comparison for SRV, GPR and LR in Fig. \ref{Fig5}. With nine Ag data points in the training set, the per-element test $R{}^2$ for Ag data already exceeds 0.5 (Table S4) for most models, and the corresponding RMSE and MAE drop to $\sim 100$ meV or lower (Fig. \ref{Fig4}c,e). At the same time, Pt predictions remain stable, with test $R{}^2$ (Table S3) and error metrics (Fig. \ref{Fig4}d,f) comparable to the Pt-only case. It is noteworthy that an outlier appeared the Pt-predictions (Fig. \ref{Fig5}a), corresponding once again to the unique $N=1$ configuration that present in the test set. Together, these results show that joint training preserves strong Pt predictive accuracy while also enabling transfer to a chemically distinct dopant. Similarly, as for the Pt-only case, training $R{}^2$ remains high ($> 0.97$) for the combined Pt+Ag dataset.

Among the models employed, GPR stands out as an exception, showing lower performance. By design, standard GPR is well suited for interpolation, but due to its zero-mean prior assumption, it tends to revert toward the training-set mean when extrapolating into a chemically distinct domain \cite{Wang2026}. Even after incorporating small number of Ag-doped configurations to the training, the models are still required to extrapolate across unsampled regions. In contrast, the other models, like linear models, extrapolate more naturally \cite{Muckley2023}, allowing them to maintain steady improvements as additional Ag-doped data are added to the training. 

Although the global RMSE and MAE of the combined Pt+Ag dataset (Fig. S10) reliably capture the improvement in predictive accuracy as more Ag-doped configurations are added to the training set, relying solely on the global test $R{}^2$ can be misleading. The models exhibit a high global test $R{}^2$ ($> 0.9$) (Table S2) even when the predictions for the Ag-doped data remain comparatively poor (Table S4). This behavior likely arises from the definition of $R{}^2$ (Eq. \eqref{R}). The larger variance in formation energies across the combined Pt+Ag dataset inflates the metric and can obscure poor performance on specific subsets. Thus, per-element metrics are essential to avoid overestimating model performance. 

Despite promising performance on the fixed test set, cross-validated $R{}^2$ appear lower, generally around 0.5, after adding nine Ag-doped configurations (Fig. S11). The result underscores the sensitivity of cross‑validation to sparsely sampled dopants, which can lead to poor Ag prediction in some folds. However, the corresponding RMSE and MAE fall in the 100-200 meV range, indicating that the absolute predictive error remains reasonable. Thus, the low CV $R{}^2$ reflects the increased complexity and limited representation of Ag-doped configurations in the combined Pt+Ag chemical space, rather than a total collapse of predictive accuracy. Although the results clearly demonstrate the models' potential to learn dopant-dependent behavior, achieving stable and reliable generalization ultimately requires a more complete and balanced representation of each chemically distinct space. Encouragingly, the steady improvement upon adding Ag-doped data suggests that models can quickly adapt once each chemical space is more adequately represented, potentially including multiple configurations at each dopant concentration, as demonstrated in the Pt-only case.

\section{Conclusions}
In summary, we constructed a dataset of doped \ch{TiO2} monolayers and predicted their formation energies using physically motivated descriptors. The ML models capture the formation energies of Pt-doped configurations with high predictive accuracy, achieving low errors on the order of meV. The addition of further data yields only a marginal impact on performance, suggesting stable learning behavior and providing a reasonable basis for subsequent extensions. Moreover, we show that models trained on a single dopant can generalize to chemically distinct elements when dopant-specific data are incorporated during training. Importantly, this does not compromise Pt predictive accuracy. The local environment, particularly the coordination number of the dopants, emerges as a key descriptor influencing the predictions. Overall, the results demonstrate that limited yet carefully curated and representative datasets can capture the dominant physical trends and enable effective transferable learning across chemical domains. Nevertheless, expanding the dataset remain essential to enhance chemical-space coverage and further enable fully reliable and broadly generalizable predictions. Beyond the present system, the findings also strengthen confidence in the framework's applicability across various dopant types and related materials, while also highlighting the need for further investigations.

\section{Data availability}
The dataset and codes used for feature analysis and prediction are available on GitHub \url{https://github.com/katasik13/DFT_ML_2D_doped_TiO2_monolayer.git}. Other data supporting the findings of this study are available from the corresponding authors upon reasonable request.

\section{Acknowledgments}
K. A. acknowledges support from the I4WORLD doctoral program of the University of Oulu, and A. A. S. D. acknowledges support from the Teollisuuden Big Data project (no. 243046901). Jijo P. Ulahannan (Government College Kasaragod,  Kannur University, Kerala,  India) is thanked for useful discussions. Computational resources were provided by the CSC \text{-} IT Center for Science, Finland. 

\section{Declaration of competing interest}
The authors declare no competing financial interests.



\end{document}


\maketitle
\beginsupplement

\clearpage

\begin{table}[]
\small
\renewcommand{\arraystretch}{1.2}
\resizebox{\columnwidth}{!}{%
\begin{tabular}{|l|l|}
\hline
\textbf{Feature} & \textbf{Description}\\ \hline
$\text{N}$  & Number of dopant atoms \\ \hline
$\text{Ti}_{-}\text{dopant}_{-}\text{Ti}_{-}\text{angle}_{-}\text{min}$ & Minimum Ti-dopant-Ti angle \\ \hline
$\text{Ti}_{-}\text{dopant}_{-}\text{Ti}_{-}\text{angle}_{-}\text{max}$ & Maximum Ti-dopant-Ti angle \\ \hline
$\text{Ti}_{-}\text{dopant}_{-}\text{Ti}_{-}\text{angle}_{-}\text{mean}$ & Average of all Ti-dopant-Ti angles  \\ \hline
$\text{Ti}_{-}\text{dopant}_{-}\text{Ti}_{-}\text{angle}_{-}\text{std}$  & Standard deviation (SD) of all Ti-dopant-Ti angles  \\ \hline
$\text{dopant}_{-}\text{Ti}_{-}\text{distance}_{-}\text{min}$  &  Minimum Ti-dopant distance (\AA) \\ \hline
$\text{dopant}_{-}\text{Ti}_{-}\text{distance}_{-}\text{max}$  &  Maximum Ti-dopant distance (\AA) \\ \hline
$\text{dopant}_{-}\text{Ti}_{-}\text{distance}_{-}\text{mean}$ & Average of all Ti-dopant distances (\AA) \\ \hline
$\text{dopant}_{-}\text{Ti}_{-}\text{distance}_{-}\text{std}$  & SD of all Ti-dopant distances  (\AA) \\ \hline
$\text{dopant}_{-}\text{dopant}_{-}\text{distance}_{-}\text{min}$  & Minimum dopant-dopant distance  (\AA)  \\ \hline
$\text{CN}_{-}\text{4Å}_{-}\text{min}$ & Minimum coordination number (CN) of a dopant atom within a 4\ \AA \ radius \\ \hline
$\text{CN}_{-}\text{4Å}_{-}\text{max}$ & Maximum CN of a dopant atom within a 4\ \AA \ radius  \\ \hline
$\text{CN}_{-}\text{4Å}_{-}\text{mean}$& Average CN of all dopant atoms within a 4\ \AA\ radius  \\ \hline
$\text{CN}_{-}\text{4Å}_{-}\text{std}$ & SD of CN of all dopant atoms within a 4\ \AA \ radius  \\ \hline
$\text{q}_{-}\text{dopant}_{-}\text{min}$ & Minimum Bader charge of dopant atoms (\textit{e}) \\ \hline
$\text{q}_{-}\text{dopant}_{-}\text{max}$ & Maximum Bader charge of dopant atoms (\textit{e})  \\ \hline
$\text{q}_{-}\text{dopant}_{-}\text{mean}$  &  Average of all Bader charges of dopant atoms (\textit{e}) \\ \hline
$\text{q}_{-}\text{dopant}_{-}\text{std}$ & SD of all Bader charge of dopant atoms (\textit{e}) \\ \hline
$\text{q}_{-}\text{Ti}_{-}\text{min}$  & Minimum Bader charge of the Ti atoms nearest to dopant atoms (\textit{e}) \\ \hline
$\text{q}_{-}\text{Ti}_{-}\text{max}$ &  Maximum Bader charge of the Ti atoms nearest to dopant atoms  (\textit{e}) \\ \hline
$\text{q}_{-}\text{Ti}_{-}\text{mean}$  &  Average of all Bader charges of the Ti atoms nearest to dopant atoms (\textit{e}) \\ \hline
$\text{q}_{-}\text{Ti}_{-}\text{std}$ &  SD of all Bader charges of the Ti atoms nearest to dopant atoms (\textit{e}) \\ \hline
$\text{d}_{-}\text{band}_{-}\text{center}_{-}\text{dopant}$  & d-band center of dopant in the energy window $[-10,\ 0]$ eV (eV)       \\ \hline
$\text{WF}_{-}\text{dopant}_{-}\text{face}$ & Work function of the dopant-doped monolayer surface (eV) \\ \hline
$\text{WF}_{-}\text{O}_{-}\text{face}$  & Work function of the undoped monolayer surface (eV) \\ \hline
$\text{E}_{-}\text{fermi}$ &  Fermi level (eV) \\ \hline
$\text{E}_{-}\text{vacuum}_{-}\text{dopant}_{-}\text{face}$  &  Vacuum level of the dopant-doped monolayer surface (eV)      \\ \hline
$\text{E}_{-}\text{vacuum}_{-}\text{O}_{-}\text{face}$ &  Vacuum level of the undoped monolayer surface  (eV)     \\ \hline
\end{tabular}%
}
\caption{List of all features, calculated within the $6 \times 5 \times 1$ supercell of the doped \ch{TiO2} monolayer. Structural features are illustrated in Fig. S1.}
\end{table}

\clearpage

\begin{figure}[h!]\centering
\includegraphics[width=0.7\linewidth]{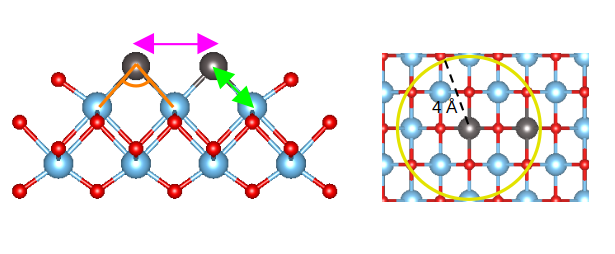}
\caption{Schematic illustration of the structural parameters used as features. On the left: green and magenta arrows denote the Ti-dopant distance and the dopant-dopant distance, respectively, and the orange arc indicates the Ti-dopant-Ti angle. On the right: the definition of CN.}
\label{fig:FigS0}
\end{figure}
\hspace{3cm}

\begin{figure}[h!]\centering
\includegraphics[width=1\linewidth]{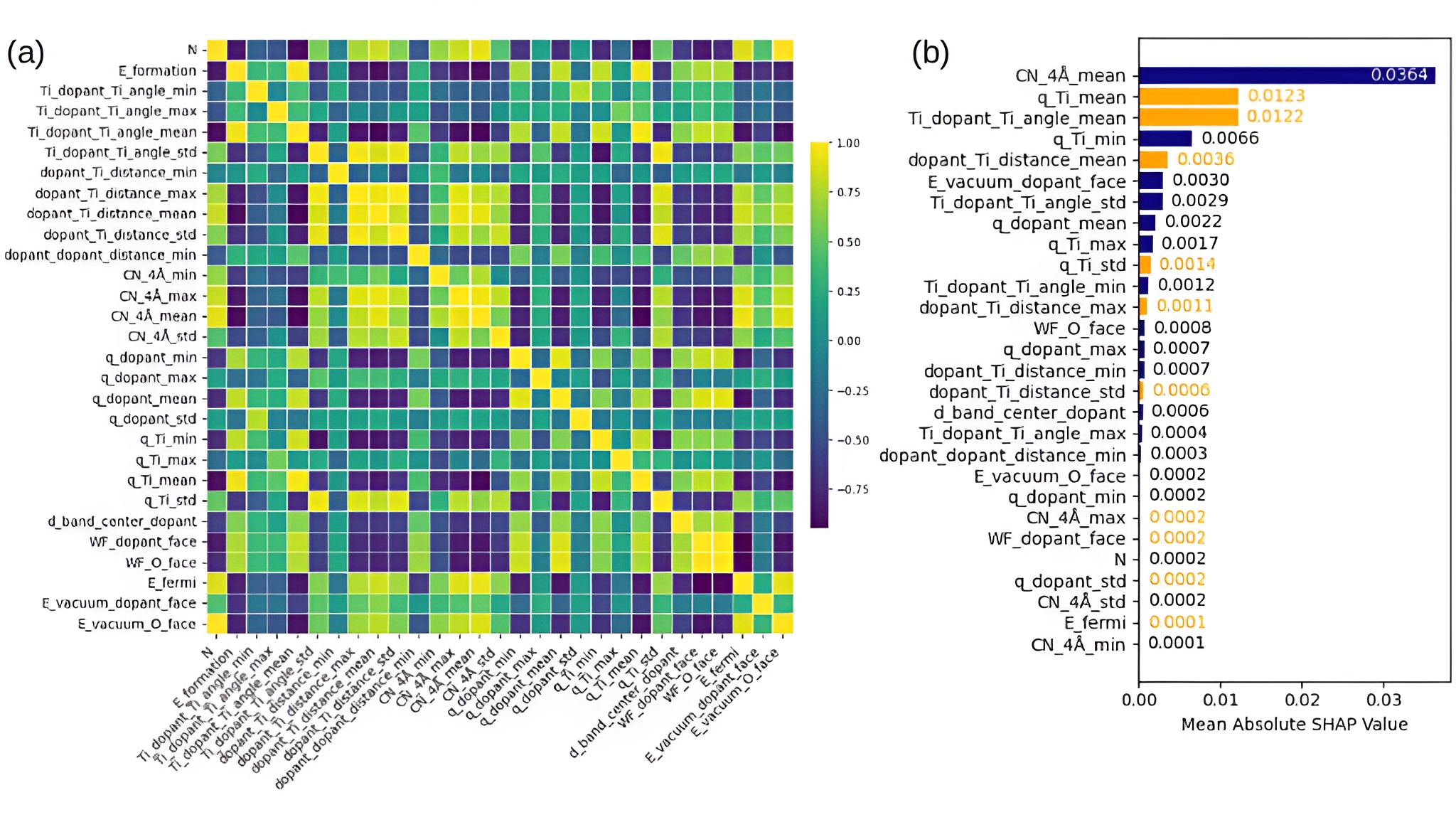}
\caption{(a) Pearson correlation coefficient matrix and (b) mean absolute SHAP value of all features, ranked by their importance. In (a), $\text{E}_{-}\text{formation}$ denotes the target formation energy per Pt atom. In (b), features removed due to multicollinearity are highlighted in orange.}
\label{fig:FigS1}
\end{figure}

\begin{figure}[!htbp]\centering
\includegraphics[width=0.75\linewidth]{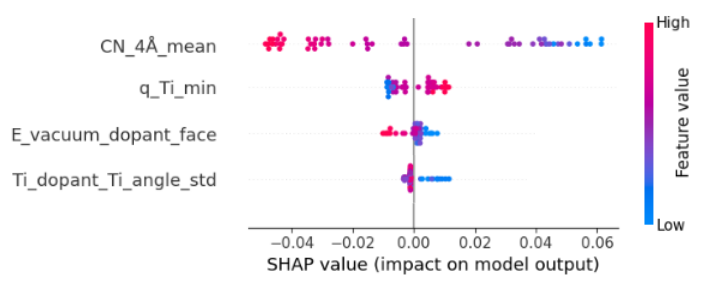}
\caption{SHAP summary plot showing the influence of the selected input features to formation-energy prediction in Pt-doped configurations. Positive (negative) SHAP value corresponds to positive (negative) impact on model output. Point colors reflects the relative magnitude of the feature value.} 
\label{fig:FigS1}
\end{figure}

\vspace{1cm}

\begin{figure}[h!]\centering
\includegraphics[width=0.95\linewidth]{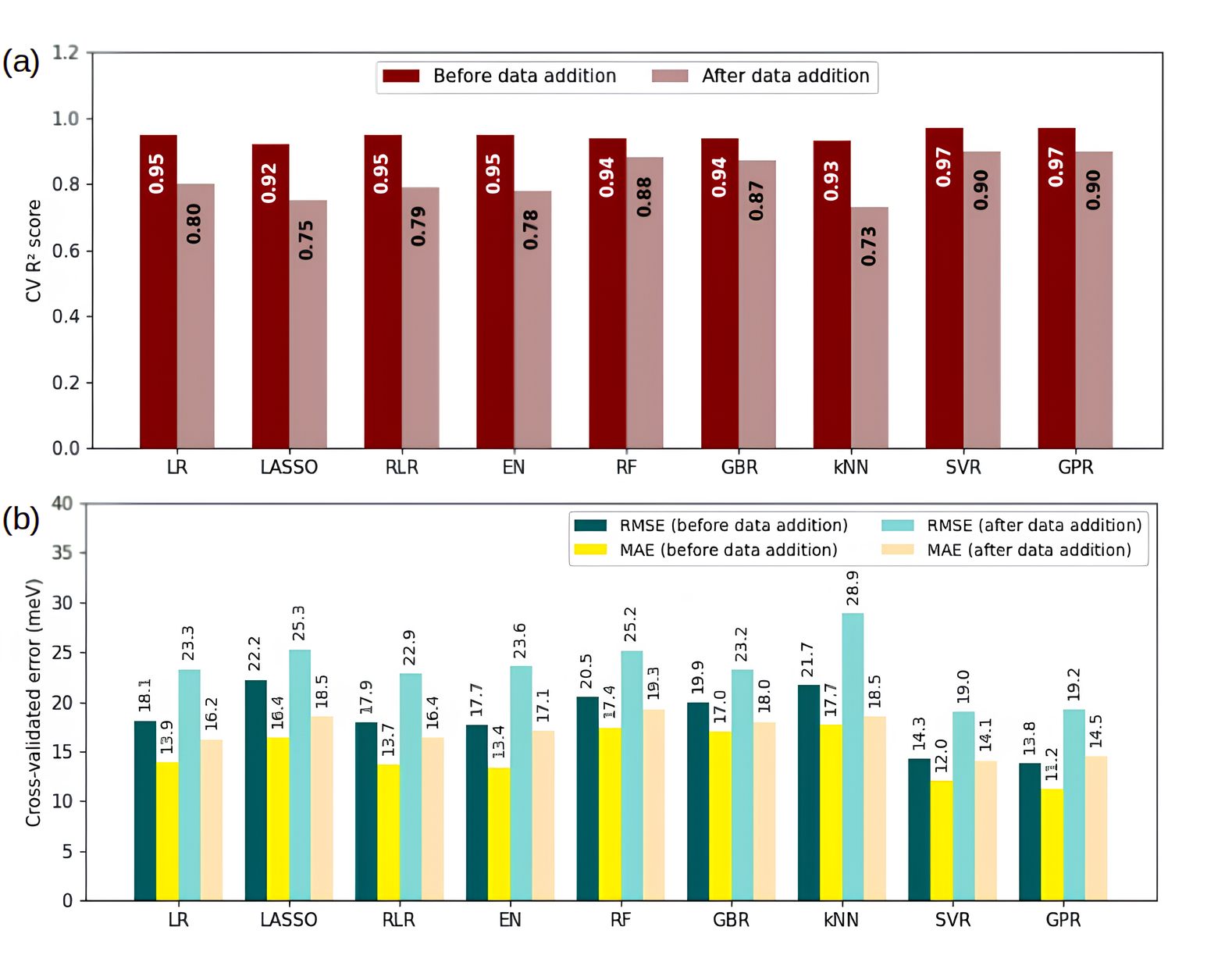}
\caption{Cross-validated (a) $R{}^2$ and (b) RMSE and MAE values of Pt-doped data before and after data addition.}
\label{fig:FigS1}
\end{figure}

\begin{figure}[h!]\centering
\includegraphics[width=1\linewidth]{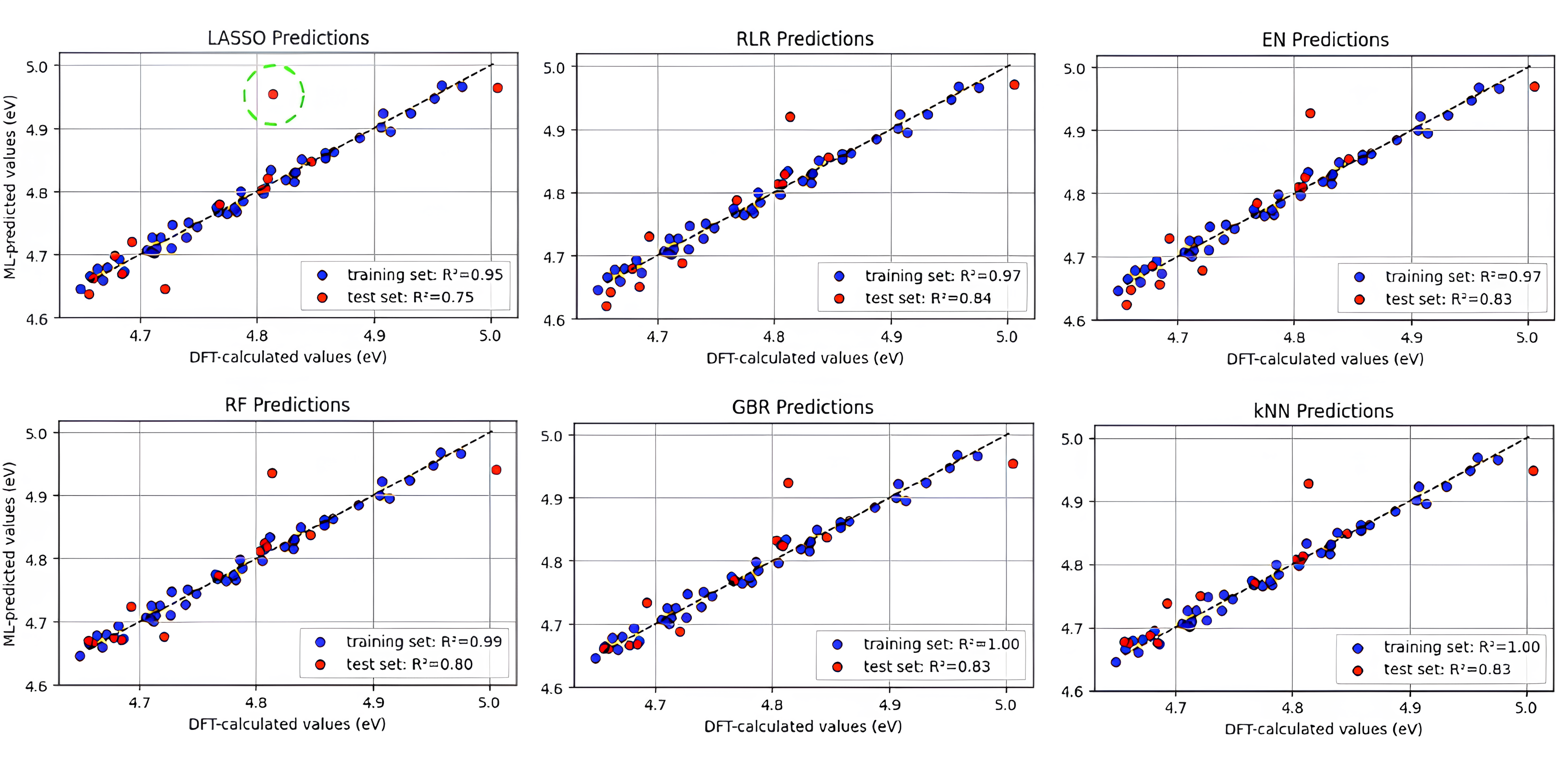}
\caption{DFT-calculated versus ML-predicted formation energy per Pt atom in eV for the training and test sets before data addition using LR, LASSO, EN, RF, GBR, and kNN models. Data point $N=1$ is highlighted with a green dashed circle in the top left plot.}
\label{fig:FigS1}
\end{figure}

\begin{figure}[h!]\centering
\includegraphics[width=1\linewidth]{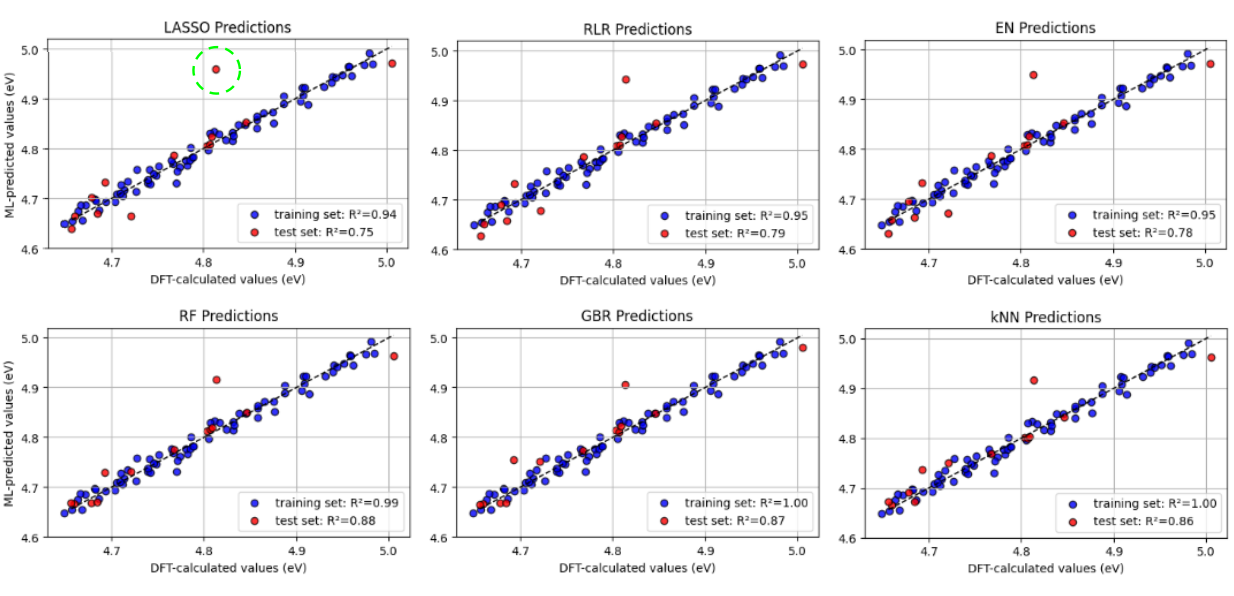}
\caption{DFT-calculated versus ML-predicted formation energy per Pt atom in eV for the training and test sets after data addition using LR, LASSO, EN, RF, GBR, and kNN models. Again, data point $N=1$ is highlighted with a green dashed circle in the top left plot.}
\label{fig:FigS1}
\end{figure}
\clearpage

\begin{figure}[h!]\centering
\includegraphics[width=1\linewidth]{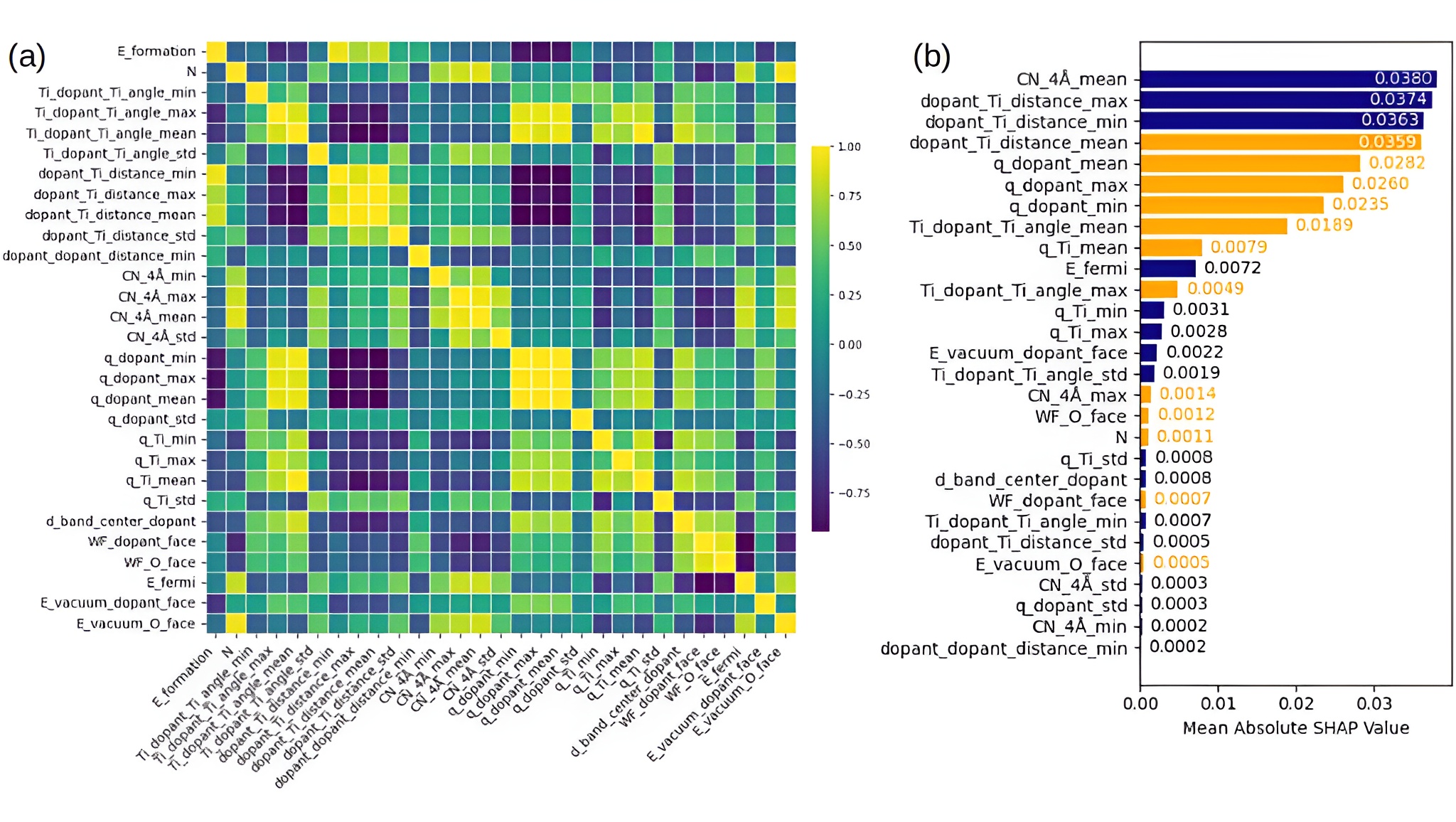}
\caption{(a) Pearson correlation coefficient matrix and (b) mean absolute SHAP value of all features after inclusion of Ag-doped data. In (a), $\text{E}_{-}\text{formation}$ denotes the target formation energy per Pt atom. In (b), features removed due to multicollinearity are highlighted in orange.}
\label{fig:FigS1}
\end{figure}

\begin{figure}[!htbp]\centering
\includegraphics[width=0.75\linewidth]{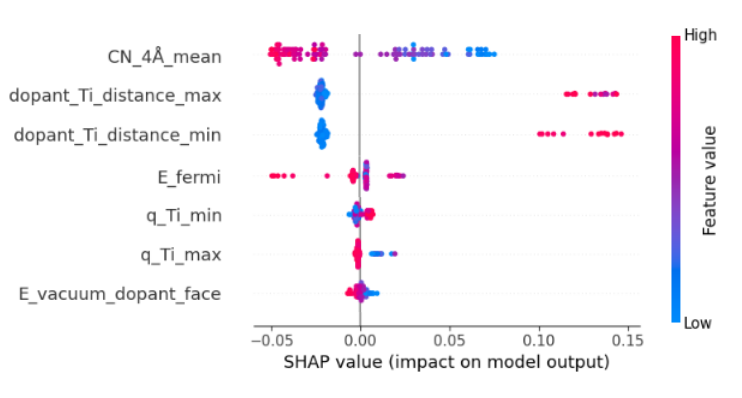}
\caption{SHAP summary plot showing the influence of the selected input features in predicting formation energies of Pt and Ag-doped configurations.} 
\label{fig:FigS11}
\end{figure}

\clearpage
\begin{figure}[h!]\centering
\includegraphics[width=1\linewidth]{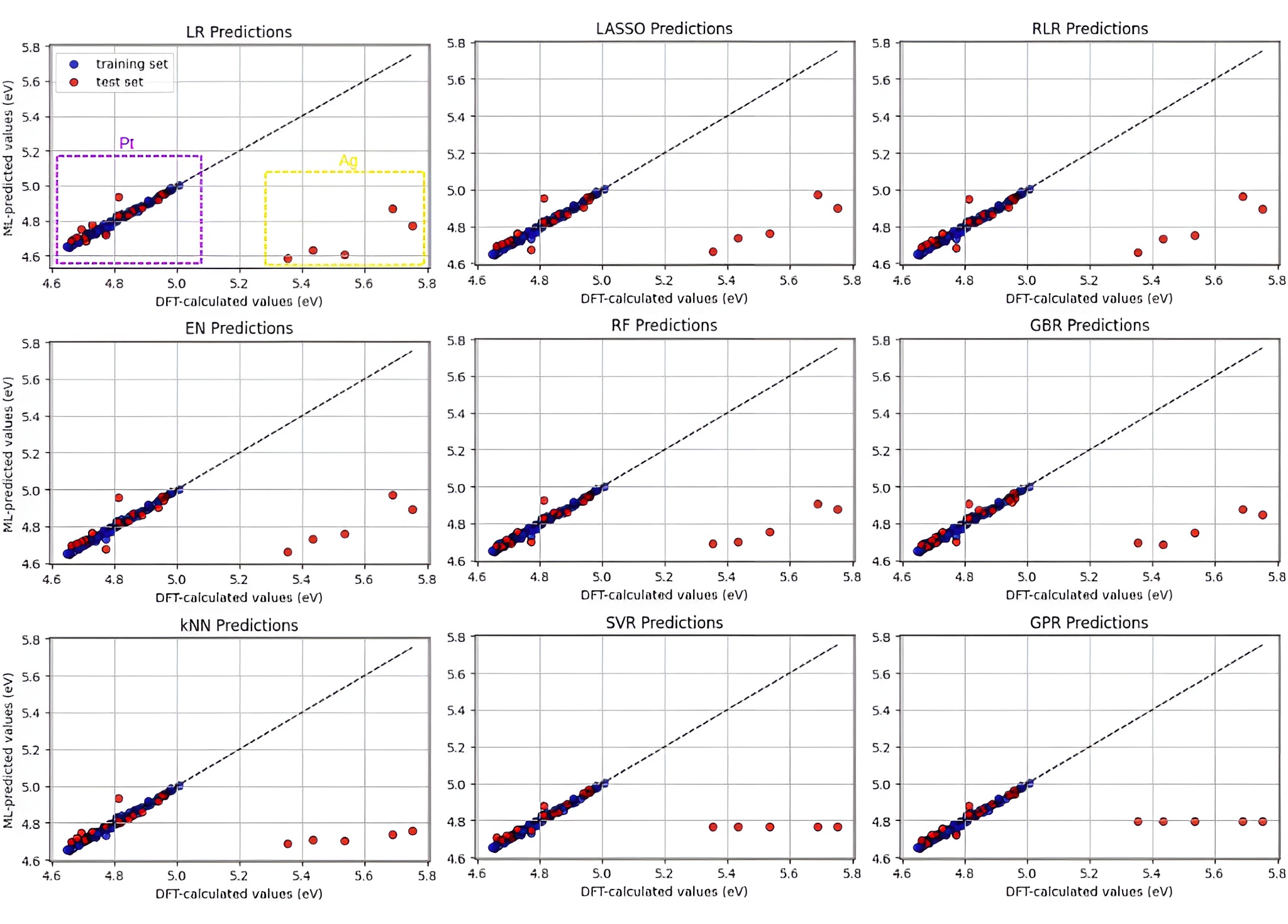}
\caption{DFT-calculated versus ML-predicted formation energy per dopant atom in eV for the combined Pt+Ag dataset without Ag-doped data added in the training. Pt and Ag-doped data points are highlighted by purple and yellow dashed squares in the top left plot, respectively.}
\label{fig:FigS1}
\end{figure}

\begin{table}[]
\begin{tabular}{llclclclc}
         & \multicolumn{7}{c}{\hspace{2.6cm}Training set composition} \\[1.05ex] \cline{3-9} 
\rule{0pt}{2.7ex}ML model &&  Pt only  && Pt + 3Ag  && Pt + 6 Ag && Pt + 9 Ag \\[1.05ex]
\hline
\rule{0pt}{2.7ex}LR &&  -0.55 && 0.80 && 0.94 && 0.97 \\
\rule{0pt}{2.7ex}LASSO && -0.17 && 0.89 && 0.95 && 0.97 \\
\rule{0pt}{2.7ex}RLR && -0.19 && 0.89 && 0.95 && 0.97 \\
\rule{0pt}{2.7ex}EN && -0.19 && 0.89 && 0.95 && 0.97 \\
\rule{0pt}{2.7ex}RF && -0.23 && 0.88 && 0.92 && 0.97 \\
\rule{0pt}{2.7ex}GBR  && -0.29 && 0.85 && 0.95 && 0.97 \\
\rule{0pt}{2.7ex}kNN && -0.49 && 0.90 && 0.92 && 0.97 \\
\rule{0pt}{2.7ex}SVR && -0.33 && 0.83 && 0.95 && 0.98 \\
\rule{0pt}{2.7ex}GPR && -0.24 && 0.89 && 0.91 && 0.85         
\end{tabular}
\caption{Global test $R{}^2$ versus the number of Ag-doped data points used for training.}\label{Table}
\end{table}

\begin{table}[!htbp]
\begin{tabular}{llclclclc}
         & \multicolumn{7}{c}{\hspace{2.6cm}Training set composition} \\[1.05ex] \cline{3-9} 
\rule{0pt}{2.7ex}ML model &&  Pt only  && Pt + 3Ag  && Pt + 6 Ag && Pt + 9 Ag \\[1.05ex]
\hline
\rule{0pt}{2.7ex}LR &&  0.85 && 0.85 && 0.88 && 0.90 \\
\rule{0pt}{2.7ex}LASSO && 0.80 && 0.87 && 0.87 && 0.85 \\
\rule{0pt}{2.7ex}RLR && 0.82 && 0.83 && 0.83 && 0.87 \\
\rule{0pt}{2.7ex}EN && 0.78 && 0.88 && 0.82 && 0.88 \\
\rule{0pt}{2.7ex}RF && 0.88 && 0.88 && 0.88 && 0.88 \\
\rule{0pt}{2.7ex}GBR  && 0.89 && 0.88 && 0.89 && 0.88 \\
\rule{0pt}{2.7ex}kNN && 0.86 && 0.90 && 0.88 && 0.90 \\
\rule{0pt}{2.7ex}SVR && 0.94 && 0.90 && 0.91 && 0.91 \\
\rule{0pt}{2.7ex}GPR && 0.94 && 0.94 && 0.94 && 0.92         
\end{tabular}
\caption{Per-element test $R{}^2$ for Pt-doped configurations versus the number of Ag-doped data points used for training.}\label{Table}
\end{table}

\clearpage
\begin{table}[ht!]
\begin{tabular}{llclclclc}
         & \multicolumn{7}{c}{\hspace{2.6cm}Training set composition} \\[1.05ex] \cline{3-9} 
\rule{0pt}{2.7ex}ML model &&  Pt only  && Pt + 3Ag  && Pt + 6 Ag && Pt + 9 Ag \\[1.05ex]
\hline
\rule{0pt}{2.7ex}LR &&  -32.06 && -3.04 && -0.13 && 0.55 \\
\rule{0pt}{2.7ex}LASSO && -23.74 && -1.14 && 0.05 && 0.50 \\
\rule{0pt}{2.7ex}RLR && -24.17 && -1.15 && 0.09 && 0.52 \\
\rule{0pt}{2.7ex}EN && -24.10 && -1.18 && 0.13 && 0.47 \\
\rule{0pt}{2.7ex}RF && -25.25 && -1.38 && -0.61 && 0.57 \\
\rule{0pt}{2.7ex}GBR  && -26.45 && -1.99 && 0.06 && 0.62 \\
\rule{0pt}{2.7ex}kNN && -30.62 && -0.98 && -0.52 && 0.55 \\
\rule{0pt}{2.7ex}SVR && -27.37 && -2.49 && 0.06 && 0.66 \\
\rule{0pt}{2.7ex}GPR && -25.45 && -1.28 && -0.83 && -2.00         
\end{tabular}
\caption{Per-element test $R{}^2$ for Ag-doped configurations versus the number of Ag-doped data points used for training.}\label{Table}
\end{table}

\begin{figure}[h!]\centering
\includegraphics[width=0.93\linewidth]{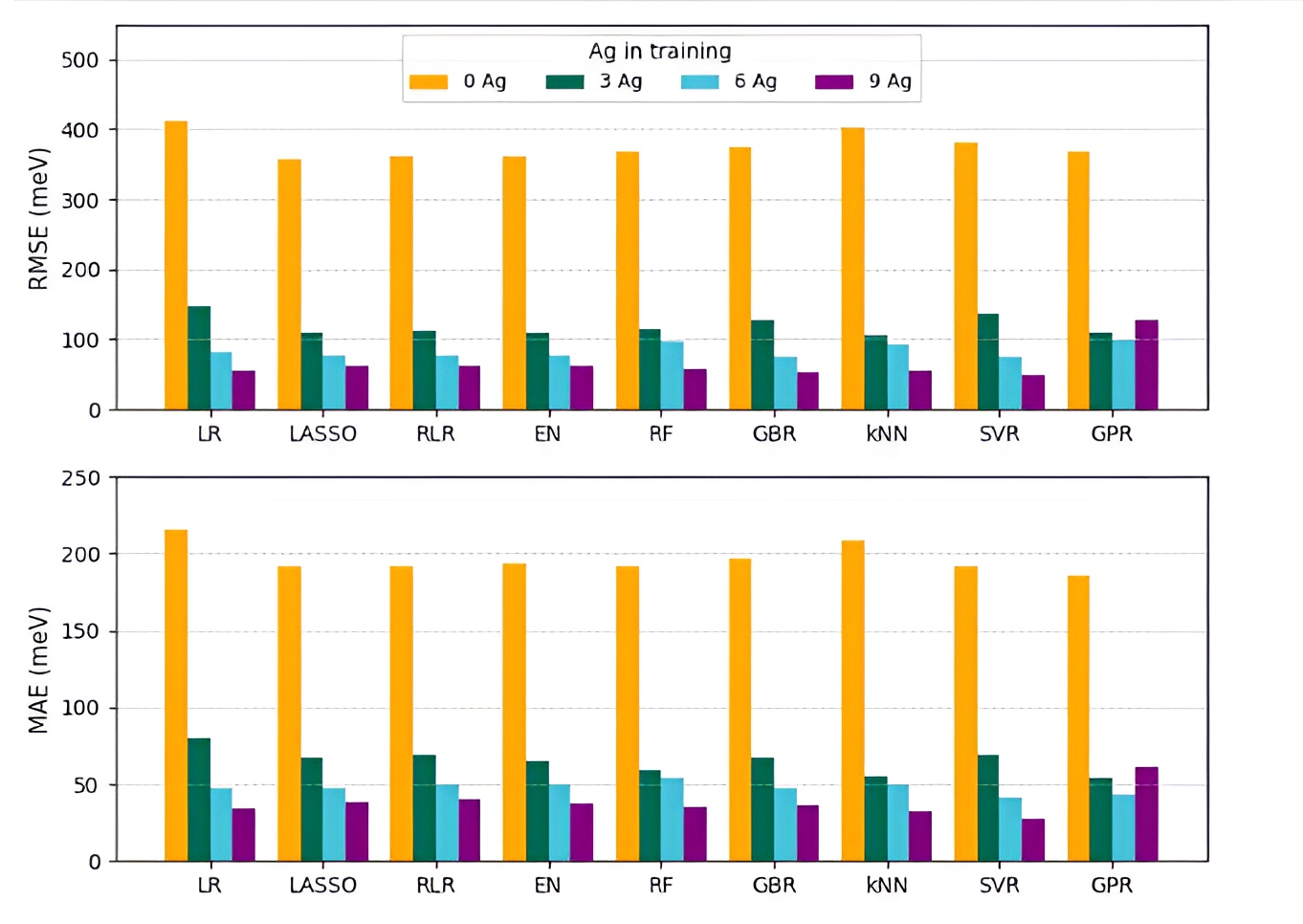}
\caption{Global RMSE and MAE for the combined Pt+Ag dataset, shown for each model, versus the number of Ag-doped data points used for training.}
\label{fig:FigS1}
\end{figure}

\begin{figure}[h!]\centering
\includegraphics[width=0.93\linewidth]{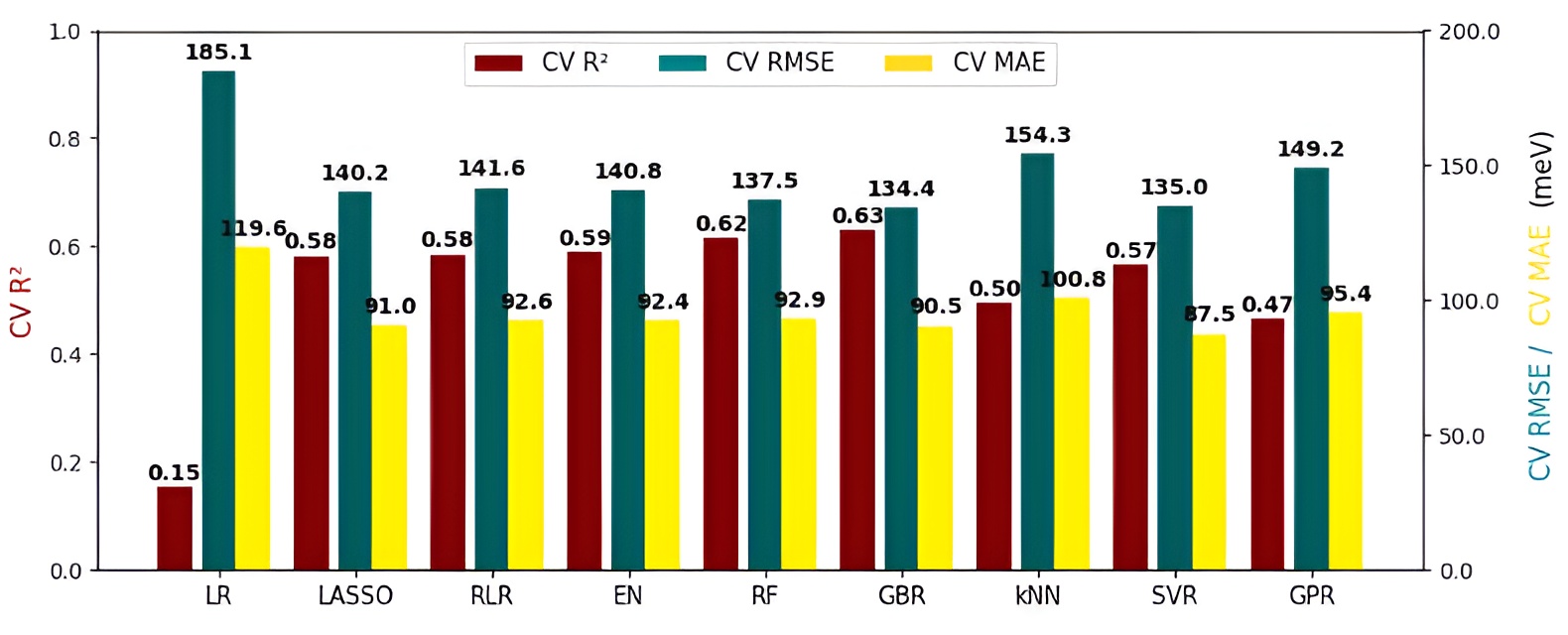}
\caption{Cross-validated results ($R{}^2$, RMSE, and MAE) of the combined Pt+Ag dataset after adding nine Ag-doped configurations to the training.}
\label{fig:FigS1}
\end{figure}
